\documentclass[aps,prl,a4paper,superscriptaddress,twocolumn,showpacs]{revtex4} 
\usepackage{graphicx} 
\usepackage{graphicx,epsfig} 
\newcommand \bea {\begin{eqnarray}} 
\newcommand \eea {\end{eqnarray}} 
\begin{document} 
\title{Non-trivial fixed point in a twofold orbitally degenerate  
Anderson impurity model } 
  
\author{Michele Fabrizio}  
\affiliation{International School for Advanced Studies (SISSA), and Instituto  
Nazionale per la Fisica della Materia (INFM) UR-Trieste SISSA, Via Beirut 2-4,  
I-34014 Trieste, Italy}  
\affiliation{The Abdus Salam International Centre for Theoretical Physics  
(ICTP),  
P.O.Box 586, I-34014 Trieste, Italy}  
\author{Andrew F. Ho}  
\affiliation{School of Physics and Astronomy,  The University of Birmingham,  
Edgbaston, Birmingham B15 2TT, UK.}  
\author{Lorenzo De Leo}   
\affiliation{International School for Advanced Studies (SISSA), and Instituto  
Nazionale per la Fisica della Materia (INFM) UR-Trieste SISSA, Via Beirut 2-4,  
I-34014 Trieste, Italy}  
\author{Giuseppe E. Santoro} 
\affiliation{International School for Advanced Studies (SISSA), and Instituto  
Nazionale per la Fisica della Materia (INFM) UR-Trieste SISSA, Via Beirut 2-4,  
I-34014 Trieste, Italy}  
\affiliation{The Abdus Salam International Centre for Theoretical Physics  
(ICTP),  
P.O.Box 586, I-34014 Trieste, Italy}  
\affiliation{INFM Democritos National Simulation Center, Via Beirut 2-4,  
I-34014 Trieste, Italy}  
\date{\today}  
\begin{abstract}  
We study a twofold orbitally degenerate  
Anderson impurity model which shows a non-trivial fixed point  
similar to that of the two-impurity Kondo model,   
but remarkably more robust, as it can only be destabilized by orbital- or  
gauge-symmetry breaking. 
The impurity model is interesting {\sl per se}, but here 
our interest is rather in the 
possibility that it might be representative of a  
strongly-correlated {\it lattice} model close to a  
Mott transition. 
We argue that this lattice model should unavoidably  
encounter the non-trivial fixed point just before the Mott transition 
and react to its instability by spontaneous generation of an orbital,  
spin-orbital or superconducting order parameter. 
\end{abstract}  
  
\maketitle  
  
When a metal is driven by strong correlations 
towards a Mott insulator, an 
incoherent component of the single-particle spectrum slowly moves 
away from the quasiparticle resonance and smoothly transforms 
into the Mott-Hubbard side-bands. Analogous behavior is displayed  
by an Anderson Impurity Model (AIM) from the  
mixed valence to the Kondo regime. 
This is suggestive of similar physical processes underlying the dynamics 
of AIM's and of strongly-correlated electron systems across the Mott metal-to-insulator 
transition (MIT), 
even though a rigorous relationship holds 
only in infinite dimensions, as shown by Dynamical Mean 
Field Theory (DMFT). 
Furthermore,   
DMFT shows that the Mott-Hubbard bands split off from the quasiparticle  
resonance quite before the MIT occurs\cite{DMFT}, 
suggesting that it is rather the physics of the AIM in the Kondo regime  
to be significant near the MIT. In that limit, the   
AIM maps onto a Kondo model with the  number of conduction   
channels always such as to {\it perfectly} screen the impurity.   
For that reason one would exclude that   
the appealing non-Fermi liquid physics of the {\it overscreened} multi-channel   
Kondo effect may ever appear close to the MIT.   
  
That expectation is partly wrong. In this Letter we study     
the phase diagram of an AIM which   
does contain a non-trivial fixed point. We show that   
the physical behavior around this fixed point resembles that   
displayed by the two-impurity Kondo model.  We also argue that   
the lattice model, where the physics of the above AIM should be relevant,   
necessarily encounters this fixed point just before the MIT,  
and discuss possible consequences.   
  
We consider the two-orbital AIM Hamiltonian:  
\begin{eqnarray}  
\hat{H}_{AIM} &=& \sum_{k,a,\sigma}\left[  
\epsilon_k \, c^\dagger_{k,a\sigma} c^{\phantom{\dagger}}_{k,a\sigma}  
+ \left( V_k c^\dagger_{k,a\sigma} d^{\phantom{\dagger}}_{a\sigma}   
+ H.c.\right)\right] \nonumber \\  
&&+ \frac{U}{2} \left(n_d -2\right)^2 + \hat{H}_K, \label{HAIM}  
\end{eqnarray}  
where $c^\dagger_{k,a\sigma}$  are conduction-band creation operators   
in orbital  
$a=1,2$ and spin $\sigma$, $d^\dagger_{a\sigma}$ impurity   
ones and $n_d = \sum_{a\sigma}   
d^\dagger_{a\sigma} d^{\phantom{\dagger}}_{a\sigma}$.  
For $\hat{H}_K=0$, the AIM has a large SU(4) symmetry which   
is lowered down to a spin SU(2) times an orbital O(2) by a Hund's rule-like coupling:  
\begin{equation}  
\hat{H}_K = K\left[  
\left(\hat{T}^x\right)^2 + \left(\hat{T}^y\right)^2\right],  
\label{HK}  
\end{equation}  
where  
$  
\hat{T}^i = \frac{1}{2}\sum_\sigma \sum_{a,b=1}^2   
d^\dagger_{a\sigma}\, \tau^i_{ab} \,  
d^{\phantom{\dagger}}_{b\sigma}  
$, $i=x,y,z$, are orbital pseudo-spin   
operators with $\tau^i$'s the Pauli matrices. For later convenience, we also introduce  
the impurity spin,   
$  
\hat{S}^i = \frac{1}{2}\sum_a \sum_{\alpha,\beta}   
d^\dagger_{a\alpha}\, \sigma^i_{\alpha\beta} \,  
d^{\phantom{\dagger}}_{a\beta} 
$,   
and the impurity spin-orbital operators 
$ 
\hat{W}^{ij} = \frac{1}{2}\sum_{a,b} \sum_{\alpha,\beta}   
d^\dagger_{a\alpha}\, \sigma^i_{\alpha\beta} \, \tau^j_{ab}  
d^{\phantom{\dagger}}_{b\beta} 
$.   
 
In the Kondo regime, 
$U$ much larger than the conduction bandwidth, two electrons   
get trapped by the impurity in a configuration identified by    
total spin, $S$, total pseudo-spin, $T$, and their  
$z$-components, with energy   
$E(S,S^z;T,T^z) = K\left[ T\left(T+1\right) - \left(T^z\right)^2\right]$.   
By Pauli principle, two electron configurations have  
either $S=1$ and $T=0$, or $S=0$ and $T=1$. 
If $K>0$, the lowest energy configuration has $S=1$ and $T=0$,    
the conventional Hund's rules. The impurity behaves   
effectively as a spin S=1 which may be Kondo-screened   
by the two conduction channels ($\delta=\pi/2$ phase shift). 
On the contrary, if $K<0$ the non-degenerate    
$S=0$ $T=1$ and $T^z=0$ state has lowest energy. Here we do not   
expect any Kondo effect, i.e. $\delta=0$. This situation is analogous to   
the two S=1/2 impurity Kondo model (2IKM) in the presence of a   
direct exchange between the impurities. There, it is known\cite{J&V,A&L,AL&J}   
that under particular circumstances an unstable fixed point (UFP)   
separates the Kondo-screened regime from the one in which the 
two-impurities couple together into a  singlet.
In our model (\ref{HAIM}), that circumstance is realized thanks to the   
O(2) orbital symmetry, as shown later, hence    
an analogous UFP should exist.  
  
We have studied the AIM (\ref{HAIM}) in the Kondo regime by    
Wilson's Numerical Renormalization Group (NRG), closely following the   
original work by Jones and Varma for the 2IKM\cite{J&V,AL&J}.    
We have also developed a complementary analysis   
based on abelian bosonization, which provides a better characterization of the UFP.       
\begin{figure}[t] 
\includegraphics[width=8cm]{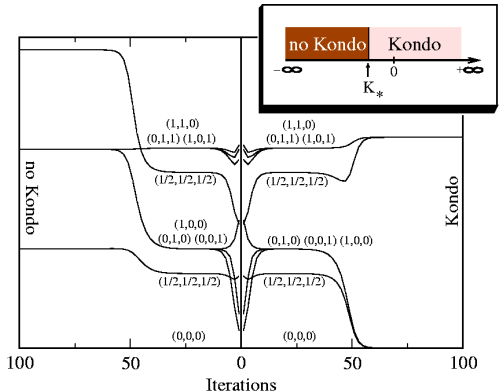} 
\caption{\label{fig1} NRG flow of the lowest energies of the levels labeled by  
$(Q,T^z,S)$, where $Q$ is half of the added charge.  
The right and left flows correspond to a relative  
deviation $\delta K_*/K_* = \pm 4\cdot 10^{-3}$ from  
the fixed point value $K_*$, respectively. The phase diagram is sketched  
in the inset.}    
\end{figure} 
 
In the inset of Fig.~1 we sketch the phase diagram of (\ref{HAIM}) in the   
Kondo regime, as determined by the flow of   
the low energy spectrum obtained by NRG. At fixed Kondo exchange    
we find indeed an UFP  
upon varying $K$.   
For $K>K_*$, the model asymptotically flows   
to a Kondo-screened fixed point, see right panel in Fig.~1, while for $K<K_*<0$ it   
flows towards a non-Kondo-screened fixed point, see left panel.   
The intermediate crossover region, also shown in Fig.~1,  
identifies the UFP\cite{units}.     
We notice that (\ref{HAIM}) has a larger impurity Hilbert space 
then the 2IKM, which contains, besides the  
$S=1$ and the $T^z=0$ $S=0$ configurations,  
also the  $T^z=\pm 1$ $S=0$ doublet, absent in the 2IKM.   
In spite of that, the low-energy spectra    
at the UFP's are the same for both models.    
\begin{figure}[t] 
\centerline{ 
\includegraphics[width=6cm]{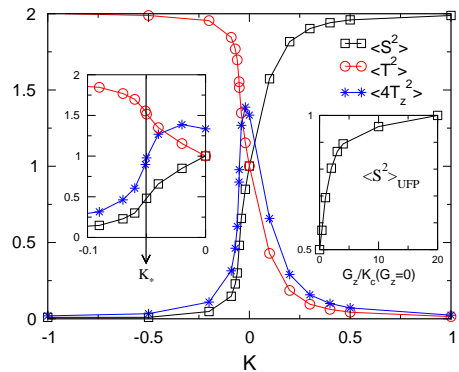}} 
\caption{\label{fig2} Average impurity quantum numbers as  
function of $K$. In the left inset the  
behavior around the UFP is shown in more detail,  
while in the right inset $\langle S^2 \rangle$ along the path  
towards the 2IKM is displayed (see text).} 
\end{figure} 
In Fig.~2 we plot the ground state average   
values of the impurity operators 
$\vec{S}^2$, $\vec{T}^2$ and $\left(T^z\right)^2$. 
For large and positive $K$, the impurity    
freezes into the $S=1$ $T=0$ configuration while, for   
very negative values, into the $T^z=0$ $S=0$ one. At the UFP  
$\langle S^2 \rangle=1/2$,   
$\langle T^2 \rangle = 3/2$ and   
$\langle \left(T^z\right)^2 \rangle = 1/4$. To prove that our UFP is 
connected with the 2IKM one,  we have added to (\ref{HAIM}) a term   
$G_z \left(\hat{T}^z\right)^2$, with $G_z>0$,   
which pushes upward the energy of the $T^z=\pm 1$ $S=0$ 
doublet absent in the 2IKM.   
In the inset of Fig.~2 we plot $\langle S^2 \rangle$   
as function of $G_z$ at the UFP, whose position depends    
on $G_z$ too. We do find that $\langle S^2 \rangle$     
smoothly reaches the 2IKM unit value for large $G_z$.

The approach to the two stable fixed points, $K<K_*$ and $K>K_*$,    
can be described by the local perturbation   
left behind by the impurity which has either disappeared, $\delta=0$, or been absorbed 
by the conduction sea, $\delta=\pi/2$: 
\begin{eqnarray}  
\hat{H}_* &=& -\sum_{a\sigma} t_*  
\left( f^\dagger_{0,a\sigma}f^{\phantom{\dagger}}_{1,a\sigma} + H.c.\right)  
+ \frac{U_*}{2}\left(n_0-2\right)^2\nonumber \\   
&& + J_{S*} \vec{S}_0\cdot \vec{S}_0 + J_{T*} \left(\hat{T}^z_{0}\right)^2,  
\label{H*}  
\end{eqnarray} 
where ``0'' labels the first available site of the Wilson chain, {\sl i.e.} the actual 
first site for $K<K_*$, $\delta=0$, and the second site for $K>K_*$, $\delta=\pi/2$.
Numerically we find    
$J_{T*}\sim 2U_* \sim 2J_{S*}\sim 32 t_* \to \infty$ upon approaching   
the UFP, see Fig.~3.  The diverging $t_*$ implies    
a singular ($\propto (K-K_*)^{-2}$) impurity-contribution to the specific heat,  
just like in the 2IKM.   
\begin{figure}[b] 
\centerline{ 
\includegraphics[width=6cm]{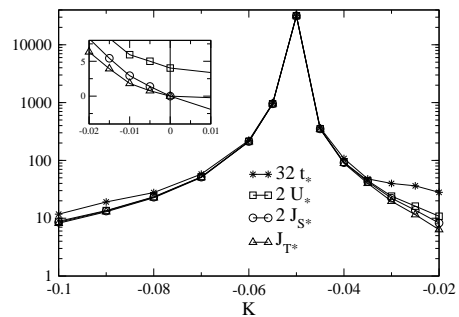}} 
\caption{\label{fig3} $K$-dependence of the effective couplings in Eq. (\ref{H*}).  
In the inset the region around $K=0$ is shown.  
} 
\end{figure} 
Additional information are provided by the Wilson ratios  
$ 
R_i = \left(c_v \, \delta \chi_i\right)/\left(\chi_i\, \delta c_v\right) 
$,  
where $i=S,T^z$ refer to spin and orbital susceptibilities, i.e. to those response   
functions related to  conserved quantities, hence accessible by   
Fermi liquid theory \cite{Nozieres,Zawadowskii}. $R_S$ and $R_{T^z}$ are shown in   
Fig.~4 and appear to vanish at the UFP. By analogy with the 2IKM, there are two other    
susceptibilities which are instead expected to be singular: 
the susceptibility $\chi_{ST^z}$ to a field which   
couples to the relative spin operator $\hat{W}^{iz}$, $i=x,y,z$, 
(the staggered spin-susceptibility in the 2IKM),  
and the pairing susceptibility $\chi_{SC}$ in the Cooper channel   
$c^\dagger_{1\uparrow}c^\dagger_{2\downarrow}   
+ c^\dagger_{2\uparrow}c^\dagger_{1\downarrow}$. Those are not accessible by    
Fermi liquid theory. Yet one can get a rough estimate of them by the   
corresponding scattering amplitudes at zero external frequencies.  
They are given respectively by $\Gamma_{ST^z}= -2U_* +J_{S*}-J_{T*}$ and   
$\Gamma_{SC}= 2U_* -3J_{S*}-J_{T*}$,   
hence are negative (corresponding to an enhancement of the response) and diverge similarly  
approaching the UFP. The physics underneath is the same of the 2IKM,   
and has been exhaustively discussed in Ref. \cite{AL&J}. The UFP has a   
residual entropy $\ln \sqrt{2}$. Away from the UFP, this entropy is   
quenched below a temperature scale $T_*\sim |K - K_*|^2$, 
implying a specific heat coefficient   
$\gamma \sim 1/T_*$. The rest of the impurity entropy     
is quenched at higher temperatures of order $T_K\sim |K|$. At the UFP,   
$\gamma$ is finite, whilst both $\chi_{ST^z}$ and $\chi_{SC}$ display   
a $|\ln T|$ singularity.   
\begin{figure}[t] 
\centerline{ 
\includegraphics[width=6.4cm]{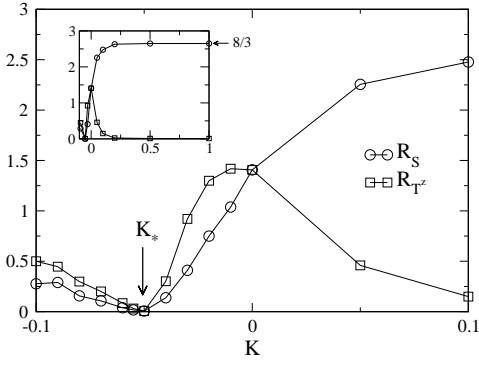}} 
\caption{\label{fig4} Spin ($R_S$) and orbital ($R_{T^z}$) Wilson ratios as functions of $K$.  
Notice that the $K=0$, SU(4)-point, as well as large $K$, $S=1$ impurity (shown in the inset),  
values coincide with known results.    
} 
\end{figure} 

The stability of the UFP is more easily accessed by abelian bosonization,  
following Ref.~\cite{Gan} on the 2IKM. In the large $U$ limit, 
(\ref{HAIM}) maps onto the Kondo model   
\begin{eqnarray}   
\hat{H}_{s-d} &=&  \sum_{k,a,\sigma}\,\epsilon_k \,  
c^\dagger_{k,a\sigma} c^{\phantom{\dagger}}_{k,a\sigma} + \hat{H}_K  
+ \sum_{i,j=1}^3 J_W^{ij} \,  \hat{W}^{i j} \, \hat{\omega}^{ij} \nonumber\\ 
&&+\sum_{i=1}^3  \left( J_{S}^i \, \hat{S}^{i} \, \hat{\sigma}^i 
 +  J_T^i \, \hat{T}^{i} \, \hat{\tau}^i\right), 
\label{Hsd} 
\end{eqnarray} 
where $\hat{\sigma}^i$, $\hat{\tau}^i$ and $\hat{\omega}^{ij}$ are, respectively, the  
conduction-electron spin, orbital and spin-orbital densities at the  
impurity site. As usual in abelian bosonization    
we allow for anisotropy: $J_S^x = J_S^y \neq J_S^z $, and similarly for $J_T^i$ and   
$J_W^{ij}$. We further assume $J_W^{ij}=J_W^\perp$, for $i,j\not = z$.
The anisotropic Kondo model (\ref{Hsd}) has   
a continuous O(2)$_{spin} \times$ O(2)$_{orbital} \times$ U(1)$_{charge}$ symmetry,   
which is useful to decompose into 
U(1)$_{spin}\times$U(1)$_{orbital}\times$U(1)$_{charge}$ plus  
two discrete symmetries: (i) a $\Pi_x^{spin}$ rotation:   
$c_{a,\sigma} \leftrightarrow c_{a,-\sigma}$  
and $d_{a,\sigma} \leftrightarrow d_{a,-\sigma}$; and (ii) a $\Pi_x^{orb}$ rotation:  
$c_{1\sigma} \leftrightarrow c_{2\sigma}$ and  
$d_{1\sigma} \leftrightarrow d_{2\sigma}$.   
By abelian bosonization\cite{ZarandDelft}, we write the $s$-wave scattering components  
of conduction electrons as chiral one-dimensional fermions    
$c_{a,\sigma}(x)=1/\sqrt{2\pi\alpha} \, F_{a,\sigma} \exp\left[-i \phi_{a,\sigma}(x)\right]$,  
where $\phi_{a,\sigma}$ are chiral free Bose fields, $\alpha$ a short distance cut-off, and  
the Klein factors $F_{a,\sigma}$ are Grassman variables enforcing proper   
anticommutation relations. Next, we introduce  
the combinations:  
$\phi_c = (\phi_{1\uparrow} + \phi_{1\downarrow} +\phi_{2\uparrow} +\phi_{2\downarrow})/2$,  
$\phi_s = (\phi_{1\uparrow} - \phi_{1\downarrow} +\phi_{2\uparrow} -\phi_{2\downarrow})/2$,  
$\phi_f = (\phi_{1\uparrow} + \phi_{1\downarrow} -\phi_{2\uparrow} -\phi_{2\downarrow})/2$,  
$\phi_{sf} = (\phi_{1\uparrow} - \phi_{1\downarrow} -\phi_{2\uparrow} +\phi_{2\downarrow})/2$.  
After applying the canonical transformation  
$\exp\left[ i \hat{S}^z \phi_s(0) \right]  \exp\left[ i \hat{T}^z \phi_f(0) \right]$,   
Eq.~(\ref{Hsd}) can be re-fermionized via   
$\Psi_b(x) = 1/\sqrt{2\pi\alpha}\, F_{b} \exp\left[-i \phi_{b}(x)\right]$,  
where $b=c,s,f,sf$\cite{ZarandDelft}.   
For a particular value of $J_S^z=J_T^z$, the end result is an effective model 
where only $\Psi_{sf}$ is coupled to the impurity, just like in  
the 2IKM\cite{Gan}.   
  
To locate the UFP, we follow the same  
strategy of Ref.~\cite{Gan}: we assume $K$ large compared to  
the conduction-bandwidth and  
search for an accidental ground state degeneracy in   
that part of the effective Hamiltonian involving just the impurity and the  
$F_b$'s:  
\bea \label{Himp}  
\hat{H}_{imp} &=&  \hat{H}_K   
+ \lambda_S \left(\hat{S}^z\right)^2 + \lambda_T \left(\hat{T}^z\right)^2 \nonumber \\  
 &+& \frac{J_W^{\perp}}{2 \pi \alpha} \left[ F_s^{\dagger} F_f^{\dagger} \; \hat{W}^{--}  
+ F_f F_s^{\dagger}\; \hat{W}^{-+} + H.c. \right]. 
\eea  
$\lambda_S$ and  $\lambda_T$ are cut-off dependent functions of $J_S^z$, 
$J_T^z$. For a specific  
$K_*<0$, we find that the impurity state 
$\mid 0\rangle \equiv \;\mid S=0,S^z=0;T=1,T^z=0\rangle$  
is degenerate with:  
\bea \label{def:1} 
&& \mid 1 \rangle  \equiv   \frac{\cos\theta }{\sqrt{2}}  
\left( \,  
F^{\phantom{\dagger}}_f \, \mid 0,0;1,+1\rangle +  
F^\dagger_f \, \mid 0,0;1,-1\rangle \,\right)  
\nonumber \\ 
&&~~ + \frac{\sin\theta }{\sqrt{2}}  
\left( \,  
F^{\phantom{\dagger}}_s \, \mid 1,+1;0,0\rangle -  
F^\dagger_s \, \mid 1,-1;0,0\rangle \,\right),  
\eea  
where $\theta$   
depends on the Hamiltonian parameters. For our model (\ref{HAIM}),  
$\theta$ should be equal to $\pi/4$ to reproduce the observed UFP average values of  
$S^2$, $T^2$, and $(T^z)^2$.   
If we added the term  
$G_z\, \left(\hat{T}^z\right)^2$, $\theta$ should increase with $G_z$, reaching the  
2IKM value of $\theta=\pi/2$ for large $G_z$. The Klein factors in  
(\ref{def:1}) show that $\mid 0 \rangle$ and $\mid 1 \rangle$   
 differs by one fermion, justifying  
the introduction of a fictitious fermion connecting that doublet:   
$f^{\dagger} \mid 0 \rangle = \; \mid 1 \rangle$. 
  
The low-energy Hamiltonian close to the UFP, $\hat{H}_{UFP}$, is then obtained by  
projection onto the above doublet-subspace. Including up to dimension $3/2$  
operators,   
\bea \label{HUFP}  
&& \hat{H}_{UFP} =  H_0 + \lambda_0    
\left[ \Psi_{sf}^{\dagger}(0) -  
\Psi^{\phantom{\dagger}}_{sf}(0) \right] \left( f^{\dagger} + f^{\phantom{\dagger}} \right)   
 \\  
& &~~ + \lambda_{1}  \partial_x \left[ \Psi_{sf}^{\dagger}(0) -  
\Psi^{\phantom{\dagger}}_{sf}(0) \right]   
\left( f^{\dagger} - f\right)  +\delta K_*  f^{\dagger} f^{\phantom{\dagger}} , \nonumber 
\eea  
with $H_0$ the free Hamiltonian for the $\Psi_{b}(x)$'s, and $\delta K_*$ the  
deviation from the fixed-point  
value $K_*$.  $\lambda_0$ and $\lambda_1$  are model dependent parameters.
As expected, 
Eq.~(\ref{HUFP}) has the same form as in the 2IKM\cite{Gan}. The UFP   
Hamiltonian [first line of Eq.~(\ref{HUFP})] is a resonant level model involving   
one Majorana fermion    
$\Psi_{sf}^{\dagger} - \Psi_{sf}$ hybridising with $f^{\dagger} + f$.  
The combination $f^{\dagger} - f$ is free and is responsible for  
the $\ln\sqrt{2}$ UFP residual entropy.  
The relevant term (dimension $1/2$) proportional to  
$\delta K_*$ describes the deviation from the UFP,  
while the $\lambda_1$-term is the leading irrelevant  
operator (dimension $3/2$). 
Other possibly relevant operators are instead not allowed by the 
symmetry properties of (\ref{Hsd}), which have to be preserved by 
$\hat{H}_{UFP}$ too.  
For instance, among the particle-hole symmetry breaking terms allowed in the 2IKM\cite{AL&J}, 
only the marginal one, which does not spoil the UFP properties, may appear in 
our model, since the relevant operator, bosonization of which is given by 
$\left[\Psi^\dagger_f(0) + 
\Psi^{\phantom{\dagger}}_f(0)\right]\left(f-f^\dagger\right)$\cite{Gan},
is here forbidden by U(1)$_{orbital}$. In fact, while 
$f$ is invariant  
under a U(1)$_{orbital}$ rotation parametrized  
by a phase $\alpha$,  
due to the Klein factors in (\ref{def:1}), $\Psi_f$ transforms into 
${\rm e}^{2i\alpha} \Psi_f$. 
Indeed all relevant perturbations which destabilize the UFP correspond to physical 
instabilities of model 
(\ref{HAIM}), unlike what happens in the 2IKM. 
For instance, the relevant terms      
$\left[ \Psi^{\phantom{\dagger}}_c(0) \pm \Psi^\dagger_c(0) \right] 
\left(f-f^\dagger\right)$,  
of dimension 1/2, break U(1)$_{charge}$. Therefore, gauge  
symmetry breaking destabilizes the UFP, which explains 
the singular behavior of $\chi_{SC}$. 
Analogously, $\chi_{ST^z}$ is the response to a   
field which breaks SU(2)$_{spin}\times \Pi_x^{orb}$
and allows the relevant terms\cite{nota}   
$ 
\left[\Psi^\dagger_{sf}(0) + \Psi^{\phantom{\dagger}}_{sf}(0)\right]\left(f-f^\dagger\right) 
$\cite{Gan,AL&J} and  
$ 
\left[\Psi^\dagger_{s}(0) \pm \Psi^{\phantom{\dagger}}_{s}(0)\right]\left(f-f^\dagger\right) 
$. Besides those two susceptibilities, also  
$\chi_{T^a}$ and $\chi_{ST^a}$, with $a=x,y$, are logarithmically diverging, being  
related to fields breaking U(1)$_{orbital}$. 

We now turn to our original motivation and discuss the possible relevance of the above   
results to the physics of the Mott transition. Take a lattice model 
with  
an on-site interaction of the   
same form as in (\ref{HAIM})-(\ref{HK}), with inverted Hund's coupling $K<0$. 
This may occur if 
the electrons are Jahn-Teller coupled to two degenerate weakly dispersive 
optical phonons by  $g\sum_R \left(q_{1R} \hat{T}^x_R 
+ q_{2R} \hat{T}^y_R\right)$, where $q_{iR}$ are the phonon coordinates on site $R$. 
This coupling gives rise to a retarded electron-electron interaction which 
reduces to (\ref{HK}) with $K\simeq - g^2/\omega_0$ when 
the typical phonon frequency $\omega_0$ is much larger than the quasiparticle bandwidth. 
Alternatively, two single-band Hubbard planes/chains  coupled by  
$J\sum_R \vec{S}_{1R}\cdot \vec{S}_{2R}$, 
where 1 and 2 refer to the two planes/chains and $J>0$, 
would also display a similar behavior. 

When $K=0$ the lattice model   
should undergo a MIT at some finite $U_c$ in the absence of nesting. 
If $|K|\ll U_c$, the physics of the metallic   
phase near the MIT should resemble that of the AIM, Eq. (\ref{HAIM}), in the   
Kondo regime. Since the width of the quasiparticle resonance, {\it i.e.}  
the effective Kondo temperature $T_K$, vanishes at the MIT, $T_K\sim U_c-U$\cite{DMFT}, 
the system is forced to enter the   
critical region around the unstable fixed point, $|K|\sim T_K$,  before the MIT occurs.  
However the instability of the AIM around the UFP towards the 
orbital O(2) or charge U(1) symmetry breakings should transform in the lattice model into 
a true bulk instability. Namely, at least within DMFT, 
we expect that the self-consistency condition which relates the impurity Green's 
function with the local Green's function of the bath enlarges the UFP into a whole 
region where the model undergoes a spontaneous symmetry breaking. This would open up 
new screening channels for those degrees of freedom which survive below $T_K$ 
down to $T_* \sim |T_K + K|^2/T_K$ ($K<0$) and are responsible of the finite entropy at 
the UFP.  If the band-structure lacks nesting or Van Hove singularities, 
orbital or spin-orbital 
instabilities are not competitive with the Cooper instability\cite{susc}.    
This suggests a superconducting region 
just before the  MIT, which would be remarkable since the bare scattering 
amplitude in the Cooper channel is $U+K$, hence repulsive for $U\sim U_c \gg |K|$.     
We believe that this phase is  
analogous to the {\sl strongly correlated superconductivity} recently 
identified by DMFT in a model for tetravalent alkali doped fullerenes\cite{Capone}. 
The latter model maps by DMFT onto a threefold degenerate AIM with inverted Hund's rules,  
mimicking a $t\otimes H$ dynamical Jahn-Teller effect.
Although different from our model (\ref{HAIM})-(\ref{HK}), 
it contains the essential physics we 
have described in this work; namely the competition between the Kondo- and 
an intra-impurity-screening mechanism.   

We acknowledge helpful discussions with E. Tosatti.  
This work has been partly supported by MIUR COFIN2001 and FIRB2002,  
and EPSRC (UK) .


\begin{thebibliography}{30}
\bibitem{units} All the results presented throughout refer to a fixed Kondo exchange 
equal to 0.05 in units of half-bandwidth, which is our unit of measure for energy.
\bibitem{susc} The Bethe-Salpeter equations for bulk susceptibilities depend 
not only upon the irreducible 
vertices, which are local in DMFT, but also on products of lattice Green's functions  
in the appropriate particle-particle (p-p) or particle-hole (p-h) channels. 
While this product in the p-p channel is always singular 
by the Cooper phenomenon, in the p-h channels singularities may arise 
only due to band structure effects.   
Therefore, even though local susceptibilities hence irreducible vertices 
are singular in p-p and p-h channels, the bulk p-p susceptibility 
is more likely the dominant one.
\end{thebibliography}
\end{document}